\documentstyle[aps,epsf]{revtex}  % DO NOT DELETE THIS LINE

\def\DEDX{d$E$/d$x$ }

\newcommand{\eepm}     {e^{+}e^{-}}

\newcommand{\rarr}     {\rightarrow}

\begin{document}
\baselineskip 14pt

\title{BES Results on Inclusive D Meson Decays}
\author{XinChou Lou}
\address{University of Texas at Dallas}

\maketitle              % Creates the title area, Do Not Remove

\begin{abstract}        % Do Not Delete this line

A measurement of branching fractions of 
the $D^0$ and $D^+$ mesons
into the $\phi$ meson is reported.
The result is based on  
a data sample of 22.3 $\rm {pb^{-1}}$
collected at the CM energy of 
4.03 GeV with the BES detector operated at
the BEPC $\rm {e^+e^-}$ storage ring.
From tagged ${\rm D\overline{D}}$ pair events
the average branching fraction for a mixture
of D$^0$ and D$^+$ is determined to be
$B(\rm{D}\rightarrow \phi \rm X)=
(1.29\pm0.51\pm0.12)\%$.
Upper limits at 90\% confidence level are set to be
$B(\rm{D^0}\rightarrow \phi \rm X)<2.5\%$, 
$B(\rm {D^+}\rightarrow \phi \rm X)<5.0\%$, and
$B(\rm {D^+}\rightarrow \phi e^+ \nu)<1.6\%.$

\end{abstract}

\section{Introduction}

In an era of high precision
experiments such as the B factories and the LHC, accurate measurements
of b-flavored particles can benefit from 
a better knowledge of charm decays and their branching
fractions.
The inclusive decay 
${\rm D}\rightarrow\phi \rm X$ 
has not been measured \footnote{Throughout this paper,
charge conjugation is implied.}.
This branching fraction can serve as an
independent check of the existence of
additional exclusive decays of D mesons that contain a $\phi$
meson\cite{pdg},
and for ${\rm B^0_s}$ physics studies
that use the $\phi\ell$ pair to tag the ${\rm B_s^0}$ 
meson\cite{bsmixing}.
In addition, this branching fraction would be helpful in understanding
the charm meson decay mechanisms.

In this paper, we report a first measurement of the inclusive $\phi$
decay branching fractions of charged and neutral $\rm D$ mesons and a new
search for the exclusive semileptonic decay
$\rm {D^+} \rarr \phi \rm {e^+} \bar{\nu}$.

\section{Data Sample and Analysis Methods}

This measurement is based on 22.3 $\rm {pb^{-1}}$ of
data collected in $\rm {e^+e^-}$ annihilations at $\sqrt{s}=4.03$ GeV
at the BEPC during the 1992-1994.
The BES detector has been described in detail elsewhere\cite{bes}.

At $\sqrt{s}$=4.03 GeV charm mesons D$^0$ and D$^+$ 
are produced via

\indent
{\hskip 1.5cm}
$\rm {e^+e^-}  \rarr \rm {D^+D^-,D^0\overline{D^0}}, $\\
\indent
{\hskip 2.9cm} $\rm {D^{+}D^{*-},D^{*+}D^{-}, D^0 \overline{D^{*0}}} $\\
\indent
{\hskip 2.9cm} $\rm {D^{*+} D^{*-},D^{*0} \overline{D^{*0}}} $\\

\noindent followed by cascade decays of the 
D$^{*}$ mesons.  
However, the D$^{*-}$ can decay either to $\pi^{-} \overline {\rm {D^0}}$
or $\pi^0(\gamma) \rm {D^{-}}$, 
so that reconstructing a D meson does not necessarily determine
whether the recoiling D meson is charged or neutral.
In order to measure specifically
$B{\rm (D^0\rightarrow \phi X)}$ and
$B{\rm (D^+\rightarrow \phi X)}$,
the numbers of neutral and 
charged D mesons recoiling against a reconstructed D meson,
and the type of the D meson from 
which the $\phi$ mesons come,
must be determined.
To this end two methods have been developed and are used
to measure the inclusive branching fractions of the D mesons.

\subsection{The $\rm {D^0}$ and $\rm {D^+}$
combinative double tag method {\rm (CDTM)}}

To measure inclusive $\phi$ branching
fractions of the ${\rm D^0}$ and ${\rm D^+}$
mesons,
the $\phi$ is searched in the recoil side against a 
fully reconstructed D meson,
and the numbers of $\phi$ events 
against the ${\rm D^0}$ and ${\rm D^+}$
decays, N$_{\rm {D}_{tag}^{0}}^{\phi}$, 
N$_{\rm {D}_{tag}^{+}}^{\phi}$, are 
determined, which can be related via

\begin{small}
\begin{equation}
{\rm N}_{\rm {D}_{tag}^{0}}^{\phi}=
\epsilon{\hskip 0.3mm} {\rm N}_{\rm {D}_{tag}^{0}}^{\rm {D^-}}
 {\hskip 0.2mm}
B({\rm D^-} \rarr {\phi}{\hskip 0.3mm}{\rm X}) +
\epsilon {\hskip 0.3mm}{\rm N}_{\rm {D}_{tag}^{0}}^{\overline {\rm {D^0}}}
{\hskip 0.2mm} B(\overline {\rm {D^0}} \rarr {\phi}{\hskip 0.3mm} {\rm X}),
\end{equation}
\end{small}

\begin{small}
\begin{equation}
{\rm N}_{\rm {D}_{tag}^{+}}^{\phi}=
\epsilon{\hskip 0.3mm} {\rm N}_{\rm {D}_{tag}^{+}}^{\rm {D^-}}
 {\hskip 0.2mm}
B({\rm D^-} \rarr {\phi}{\hskip 0.3mm}{\rm X}) +
\epsilon {\hskip 0.3mm}{\rm N}_{\rm {D}_{tag}^{+}}^{\overline {\rm {D^0}}}
{\hskip 0.2mm} B(\overline {\rm {D^0}} \rarr {\phi}{\hskip 0.3mm} {\rm X}),
\end{equation}
\end{small}

\noindent
to the branching fractions of their decays, 
B(${\rm D^- \rarr \phi X}$) and B(${\rm D^0 \rarr \phi X}$),
where ${\rm N_{D_{tag}^0}^{D^-}}$,  
${\rm N_{D_{tag}^0}^{\overline{D^0}}}$, 
${\rm N_{D_{tag}^+}^{D^-}}$, and 
${\rm N_{D_{tag}^+}^{\overline{D^0}}}$ 
are respectively
the numbers of ${\rm D^-}$ and $\overline {\rm D^0}$
decays on the recoil against 
$\rm D^+$ and ${\rm D^0}$ tags, 
and $\epsilon$ is the detection
efficiency of the $\phi$.
The values of ${\rm N_{D_{tag}^0}^{D^-}}$, 
${\rm N_{D_{tag}^0}^{\overline{D^0}}}$,
${\rm N_{D_{tag}^+}^{D^-}}$, and
${\rm N_{D_{tag}^+}^{\overline{D^0}}}$
are determined 
from a measurement of the total production
cross-sections of ractions ${\rm e^+e^-\rightarrow
D^* \overline{D^*},D^* \overline{D}}$ 
at 4.03 GeV by BES\cite{ddx}.

\subsection{The recoil charge method}

At $\sqrt{s}=$4.03 GeV, ${\rm D^* \overline{D^*}}$ and
${\rm D^*\overline{D}}$ pairs 
are produced with no additional charged tracks.
Charged pions arising from direct
D$^*$ decays are very slow, and are
mostly undetected in the BES detector. As a result, only
decay products of
the ${\rm D^+}$ and ${\rm D^0}$ are visible for most events.
Let Q$_{\rm D}$ be the charm flavor
of the reconstructed D meson, and Q$_{rec}$ be the
total charge of tracks recoiling against this D meson.
The Q$_{rec}$ distribution for D$^0$ (D$^+$) centers at 
0 (1), and has a
spread of $\pm$1.
The recoil charge method selects
neutral and charged D mesons according to

\begin{equation}
{ Q_{rec}=0, ~~{\rm or}~~Q_{rec}=Q_{\rm D}=-1 ~{\rm for}~~{\rm D^0}~tags}
\end{equation}

\noindent and
\begin{equation}
{Q_{rec}\cdot Q_{\rm D}<0~~{\rm for}~~ {\rm D^+}~tags}
\end{equation}
\noindent
For inclusive
D decays, the efficiency and the misidentification rate are 
0.74$\pm$0.02 and 0.25$\pm$0.02, respectively, 
as obtained from Monte Carlo simulations, and 
are approximately the same for both charged and neutral D mesons.
These numbers are confirmed using
kinematically selected data events 
${\rm e^+e^-\rightarrow D^+D^-}$
and ${\rm e^+e^-\rightarrow D^0\overline {D^0}}$.
For events in which 
a D tag and a recoil $\phi$ has been fully reconstructed,
the efficiency of the recoil charge method is improved over
that of the inclusive D events. A Monte Carlo study of various
D decay modes into final states containing
a $\phi$ has been
performed, and the variations among their efficiencies
are included in the systematic errors. 
For these events, the recoil charge method 
selects D meson type correctly 
0.91$\pm$0.01$\pm$0.02 of the time,
and misidentifies a D for 
0.09$\pm$0.01$\pm$0.02 of the events,
where the first error
is due to Monte Carlo statistics, and the second is 
systematic.

\section{Data Analysis}

\subsection{Reconstruction of D, $\phi$ Mesons}

Charged tracks are required to have good
helix fits which have a normalized chi-square
of less than 9 per degree of freedom.
These tracks must satisfy $|\cos\theta|<$ 0.8,
where $\theta$ is the polar angle, 
and be consistent with coming from the primary
event vertex.
For charged particles, a particle
identification procedure is applied. 
A combined particle confidence level 
calculated using the \DEDX  and TOF 
measurements is required to be greater
than $1\%$ for the $\pi$ hypothesis.
For the kaon hypothesis, ${\it L_k>L_{\pi}}$,
where ${\it L}$ is the likelihood for a 
particle type, is required.

Charged and neutral D mesons are reconstructed via
decays 
${\rm D^0 \rarr K^{-}\pi^{+}}$,
${\rm K^{-}\pi^{-}\pi^{+}\pi^{+}}$ and
${\rm D^{+} \rarr K^{-}\pi^{+}\pi^{+}}$.
To reduce combinatorial backgrounds, 
only D mesons from 
${\rm \eepm \rarr \overline{D}D^*, D^*\overline{D^*}}$
reactions are selected with cuts on the momenta of 
Kn$\pi$ combinations.
Figures 1(a), 1(b) and 1(c) show the invariant mass
distributions for events that pass the selections.
The signals are fitted, and after having accounted for
double counting, the number of D events is determined
to be $9054\pm309\pm416$, where the first error is
statistical and the second systematic.
These D events are used as tagged  
${\rm \eepm \rarr \overline{D}D^*, D^*\overline{D^*}}$
events in which the recoil side contain an unbiased
${\rm \overline{D}}$ decay.

\begin{figure}[h]      % in second brace, h=here, t=top, b=bottom
\centerline{\epsfxsize 5.0 truein \epsfbox{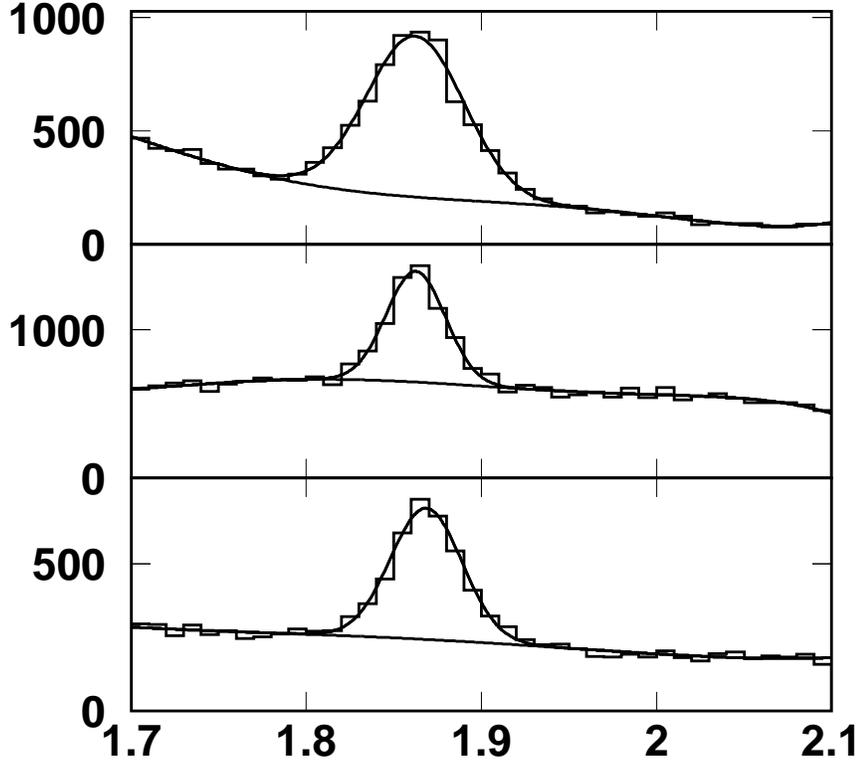}}
\vskip -.2 cm
\caption[]{
\label{Figure 1}
\small Invariant mass distributions for ${\rm K^-\pi^+}$ (top),
${\rm K^-\pi^+\pi^+}$ (middle), and 
${\rm K^-\pi^+\pi^+\pi^-}$ (bottom).}
\end{figure}

Table 1 summarizes the numbers of neutral and charged D mesons
in the recoil against the reconstructed D tags. The averages
from the CDTM method
and the recoil charge method,
calculated assuming a full correlation between their
statistical errors,  are
6803$\pm$303$\pm$322  
and  2251$\pm$77$\pm$112 for D$^0$ and D$^+$,
respectively.

The $\phi$ meson is reconstructed 
through its decay to ${\rm K^+K^-}$.
Figure 2 shows the invariant mass distribution of ${\rm K^+K^-}$
pairs selected.
Using convoluted Breit-Wigner and Gaussian functions plus a third order
polynomial background to fit the mass spectrum, 
a mass of $1.0194\pm0.0002$ GeV$/c^2$ and a total of 
$1108\pm 70$ $\phi$ events
are obtained. In this measurement, a $\phi$ signal window
is defined as the region from 1.00 to 1.04 GeV/$c^2$, 
as indicated by the arrows
in Figure 2. 

\begin{table}
\caption{Numbers of neutral and charged D mesons on the recoil}
\begin{tabular}{rll}
 method& number of     & number of \\
       & D$^0$  events & D$^+$  events \\
 CDTM  & 6839$\pm$308  & 2215$\pm$70   \\
 recoil charge& 6767$\pm$297    & 2287$\pm$83\\
\hline
Average & 6803$\pm$303    & 2251$\pm$77   \\
\end{tabular}
\end{table}

\begin{figure}[ht]      % in second brace, h=here, t=top, b=bottom
\centerline{\epsfxsize 5.0 truein \epsfbox{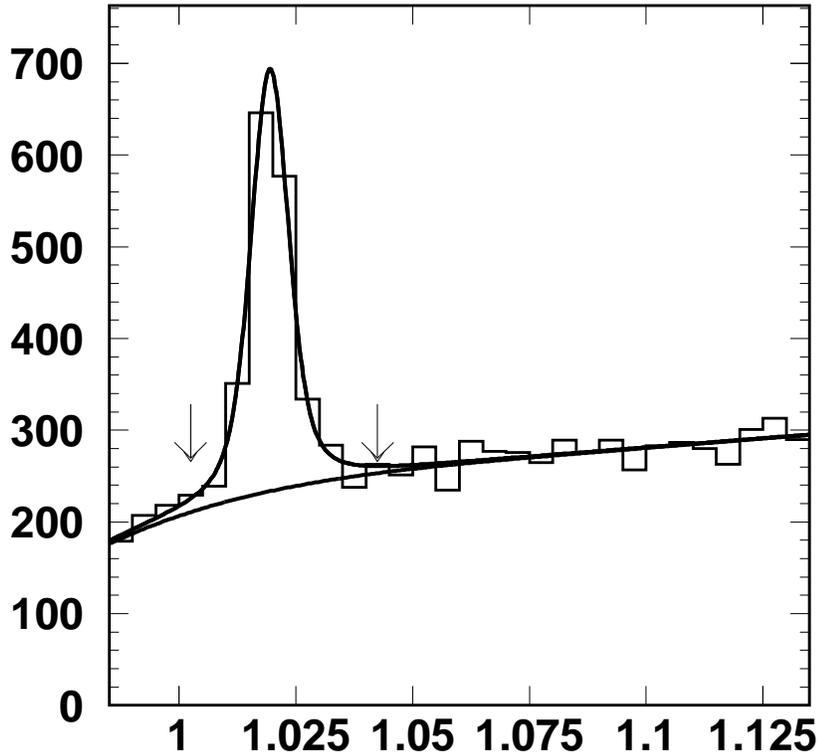}}
\vskip -.2 cm
\caption[]{
\label{Figure 2}
\small Invariant mass distribution of ${\rm K^+K^-}$ pairs.}
\end{figure}

\subsection{Inclusive ${\rm D\rightarrow \phi X}$ }

Figures 3(a) and 3(b) show 
the invariant mass distributions of ${\rm K^+K^-}$ pairs
from D$^+$ and D$^0$, respectively, as identified by the
recoil charge selection criteria.
The ${\rm Kn\pi}$ invariant masses for the single tag
are within $\pm2.5\sigma_{\rm M_{D}}$ of the $\rm D$ masses.
In this measurement, ${\rm K^+K^-}$ pairs with
masses in the ranges  
0.98 - 1.00 GeV/$c^2$ and 1.04 - 1.15 GeV/$c^2$ are
taken as background for the $\phi$.
The ${\rm Kn\pi}$ mass
regions from 1.7 to 2.1 GeV/$c^2$, excluding
regions within ${\rm \pm3\sigma_{M_D}}$ of the fit
D masses,
are defined as background control regions for the
D mesons. As shown in Figures 3(a) and 3(b), 15 events are found as
${\rm D\phi}$ candidates, and 14 events are selected
as background outside the $\phi$ mass region.
Using the D
sideband events, a total of 0.5$\pm$0.5 background events
has been estimated as the background among the D candidates.
Subtracting the background contributions to both the D
and the $\phi$, we obtain an excess of 10.2$\pm$4.0 events in the 
$\phi$ signal region.

The two D type identification methods, CDTM and the recoil charge
method, are applied to these events to
extract the numbers of $\phi$ from specific
D$^0$ and D$^+$ decays.
Subtracting backgrounds estimated using the $\phi$ and D side
bands, the two methods determine
3.7$\pm$4.7 (CDTM) and  9.7$\pm$4.2 (recoil charge)
${\rm D^0\rightarrow\phi X}$ events, and
6.5$\pm$5.5 (CDTM), and 0.5$\pm$1.7 (recoil charge)
${\rm D^+\rightarrow\phi X}$ events, respectively.
Averaging over the two methods and assuming a complete correlation
in their statistical errors, the number of
${\rm D^0\rightarrow\phi X}$ and ${\rm D^+\rightarrow\phi X}$
events are set to be
6.7$\pm$4.5, and 3.5$\pm$3.6, respectively,
and are used to determine their branching fractions.

\begin{figure}[h]      % in second brace, h=here, t=top, b=bottom
\centerline{\epsfxsize 5.0 truein \epsfbox{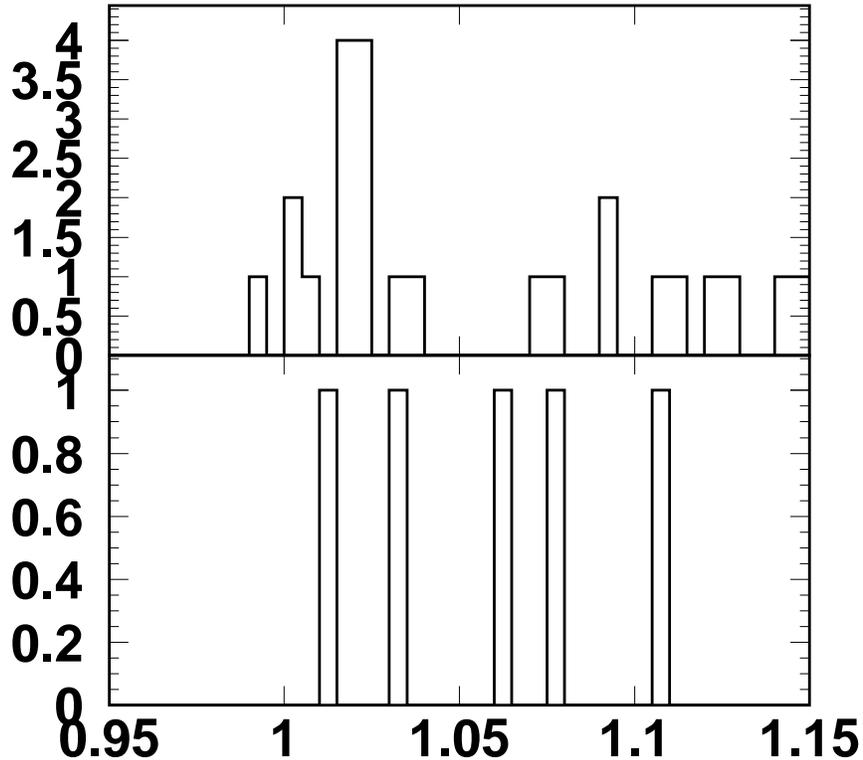}}
\vskip -.2 cm
\caption[]{
\label{Figure 3}
\small ${\rm K^+K^-}$ invariant mass distributions for 
${\rm D^0\rightarrow \phi X}$ (top), and
${\rm D^+\rightarrow \phi X}$ (bottom).}
\end{figure}

\subsection{Search for the decay $\rm {D^+}\rarr \phi e^+ \nu$} 

Among the 15 $\phi$ candidates observed in the recoil 
side of the events, 
4 are accompanied by at least one charged track which 
are within $|cos\theta|<0.85$.
Each of these tracks is checked
for consistency with being an electron 
using
the \DEDX information. This electron identification
requires that   
electron confidence level to be 
greater than $1\%$, and ${\rm L_e>L_{\pi}}$.
None of the accompanying tracks is identified as an electron.

\section{Results}

Assuming 10.2$\pm$4.0 signal ${\rm D\rightarrow\phi X}$ 
events, and correcting
for $\phi$ meson detection efficiency of 0.084$\pm$0.006 
obtained from a Monte Carlo simulation, 
the average
branching fraction for the BES mixture of D$^0$ and D$^+$
mesons is measured to be 	

$$B(\rm {D} \rarr \phi X) = (1.29 \pm 0.51 \pm 0.12)\%,$$

\noindent
where the first errors are statistical and second systematic.

Based on 6.7$\pm$4.5 ${\rm D^0\rightarrow\phi X}$ and
3.5$\pm$3.6 ${\rm D^+\rightarrow\phi X}$
events, as determined in the previous section, 
90\% C. L. upper limits are set on 
specific D$^0$, D$^+$ decays to be

$$B(\rm {D^0} \rarr \phi X) < 2.5\%,$$

$$B(\rm {D^+} \rarr \phi X) < 5.0 \% $$

\noindent
The results include systematic errors arising 
from uncertainties
($\pm0.05\%$, $\pm0.06\%$ and $\pm0.04\%$)
in the numbers of singly tagged D mesons due to
the choice of a background function and fit interval for the single
tag samples and 
uncertainties ($\pm0.08\%$, $\pm0.13\%$ and $\pm0.09\%$)
in the inclusive $\phi$ efficiency.
The combined effect of these sources is obtained by adding the
uncertainties in quadrature, which yields total systematic errors of
$\pm0.10\%$, $\pm0.14\%$ and $\pm0.10\%$
for the D$^0$, D$^+$, and their sum,  respectively.

Based on zero candidate ${\rm D^+\rarr \phi e^+ \nu}$ events,
and a detection efficiency of 0.0652,
a 90\% C. L. limit is set for the decays at

$$B(\rm {D^+} \rarr \phi e^+ X) < 1.6 \% .$$

\section{Conclusion}

In summary, the inclusive 
branching fractions of the charged and
neutral D mesons into a $\phi$ 
have been directly measured. 
Comparing with the sums of the existing measurements
on the exclusive $\rm D^0$ and $\rm D^+$ decays
containing a $\phi$ in the final states,
these BES branching fraction values indicate
little room for additional $\phi$ decay modes
of D$^0$ and D$^+$ mesons.

\end{document}